\documentclass[
floatfix, 
twocolumn, 
showpacs, 
showkeys, 
preprintnumbers, 
nofootinbib, 
superscriptaddress
]{revtex4-1}

\usepackage[utf8]{inputenc}
\usepackage[sort&compress]{natbib}
\usepackage{ulem}
\usepackage{bm}
\usepackage{times}
\usepackage{amssymb,amsbsy,amsmath,amsfonts}
\usepackage{graphicx}
\usepackage{float}
\usepackage{color}
\usepackage{morefloats}
\usepackage{rotating}
\usepackage{srcltx}
\usepackage{slashed}
\usepackage{subfigure}
\usepackage{multirow}
\usepackage{verbatim}
\usepackage{hyperref}
\usepackage{tabularx}
\usepackage{braket}

\DeclareUnicodeCharacter{3000}{HEREHEREHERE}

\begin{document}

\title{Saturation of nuclear matter in the relativistic Brueckner Hatree-Fock approach with a leading order covariant chiral nuclear force}

\author{Wei-Jiang Zou}
\affiliation{State Key Laboratory of Nuclear Physics and Technology, School of Physics, Peking University, Beijing 100871, China}
\author{Jun-Xu Lu}
\affiliation{School of Space and Environment, Beihang University, Beijing 102206, China}
\affiliation{School of Physics, Beihang University, Beijing 102206, China}
\author{Peng-Wei Zhao}
\affiliation{State Key Laboratory of Nuclear Physics and Technology, School of Physics, Peking University, Beijing 100871, China}
\author{Li-Sheng Geng}
\email{lisheng.geng@buaa.edu.cn}
\affiliation{School of Physics, Beihang University, Beijing 102206, China}
\affiliation{Beijing Key Laboratory of Advanced Nuclear Materials and Physics, Beihang University, Beijing 102206, China}
\affiliation{Peng Huanwu Collaborative Center for Research and Education, Beihang University, Beijing 100191, China}
\affiliation{Southern Center for Nuclear-Science Theory (SCNT), Institute of Modern Physics, Chinese Academy of Sciences, Huizhou 516000, China}

\author{Jie Meng}
 \email{mengj@pku.edu.cn}
\affiliation{State Key Laboratory of Nuclear Physics and Technology, School of Physics, Peking University, Beijing 100871, China}

\date{\today}

\begin{abstract}

Nuclear saturation is a crucial feature in nuclear physics that plays a fundamental role in understanding various nuclear phenomena, ranging from properties of finite nuclei to those of neutron stars. 
However, a proper description of nuclear saturation is highly nontrivial in modern nonrelativistic~\textit{ab initio}~studies because of the elusive three-body forces.  
In this letter, we calculate the equation of state for nuclear matter in the relativistic Brueckner-Hartree-Fock (RBHF) framework with the leading order covariant chiral nuclear force. We show that a simultaneous description of the nucleon-nucleon scattering data and the saturation of the symmetric nuclear matter can be achieved. In this regard, the relativistic effects nicely explain the saturation of nuclear matter. As a result, the present study based on the covariant chiral nuclear force shows that in the RBHF framework, one can achieve saturation with a leading order covariant chiral nuclear force with only two-body forces, in contrast to the vast majorities of studies in the non-relativistic framework, where the next-to-next-to-leading order two-body and three-body chiral forces are needed. This study sets the foundation for studying nuclear saturation with the covariant chiral force in the RBHF framework, which allows for a systematic understanding of one of the key features of nuclear physics more microscopically.
\end{abstract}

\maketitle


\section{Introduction}


Nuclear matter is an ideal nuclear system of uniform density consisting of infinite neutrons and protons. The saturation of symmetric nuclear matter (SNM), i.e., the energy per nucleon reaches a minimum of about~$-16$~MeV around a density of 0.16~$\mathrm{fm}^{-3}$~neglecting the Coulomb interaction, plays a vital role in our understanding of nuclear physics~\cite{Day:1978zz,Bethe:1971xm,Lattimer:2000kb,Sorensen:2023zkk}. The historical success of some of the early phenomenological models, such as the semi-empirical mass formula~\cite{Weizsacker:1935bkz,Myers:1966zz}~and the liquid drop model~\cite{GGam30}, lies in the fact that a nucleus can be treated as an incompressible quantum liquid drop with a constant density~$\rho_0$~at first approximation. This fact alone tells that any microscopic theory of atomic nuclei should be able to describe nuclear matter satisfactorily~\cite{Machleidt:2023jws}. 

Given that the volume term is the most important one in the liquid drop model of nuclear binding energies, one may argue that saturation is a bulk property and theoretically should be independent of fine details of the nucleon-nucleon interaction. Nevertheless,  a proper description of the empirical saturation properties of SNM remains a highly nontrivial task in modern non-relativistic \textit{ab initio}~approaches. That is, one needs to supplement the two-body~\textit{NN}~forces~(2NF)~with the three-body~\textit{NNN}~forces~(3NF)~to reasonably describe the saturation of SNM~\cite{Drischler:2017wtt,Li:2008bp,Hagen:2013nca,Ekstrom:2015rta,Sammarruca:2014zia,Sammarruca:2021bpn,Lonardoni:2019ypg}.
In non-relativistic theories, the repulsion required for nuclear matter saturation is provided by many-body forces such as 3NF, which can be understood as the leading-order (in powers of $\rho_B$)
term of the velocity dependence inherent in the Lorentz scalar interaction~\cite{Forest:1995sg}. We note that
the development of consistent microscopic 3NF for the two extreme systems, finite nuclei and nuclear matter, remains a matter of current research~\cite{Hebeler:2020ocj}. More recently, many~\textit{ab initio}~calculations of nuclear structure employing chiral two- and three-nucleon interactions~(at N$^2$LO and N$^3$LO)~found that the bulk properties of medium-mass nuclei and nuclear matter saturation cannot be consistently described~\cite{Hoppe:2019uyw,Huther:2019ont,Machleidt:2023jws}. This prompted the development of interactions that use nuclear structure data of medium-mass nuclei or the saturation properties of SNM to constrain the nucleon-nucleon interaction. One such example is the~N${}^2$LO${}_{\mathrm{SAT}}$~interaction~\cite{Ekstrom:2015rta}~that simultaneously fitted the ground-state energies and radii of oxygen isotopes to obtain an~\textit{NN}+3\textit{N}~interaction at N$^2$LO. Though very successful phenomenologically, such efforts seem to deviate from the conventional definition of~\textit{ab initio}~studies (using the nuclear force calibrated with the nucleon-nucleon scattering and few-body data, and uniquely and correctly predicting the properties of nuclear matter and the structure of finite nuclei)~\cite{Entem:2002wsy,Epelbaum:2004fk,Machleidt:2023jws}. We note that how to define an ab initio theory is still a topic under debate; see, e.g., Ref.~\cite{Ekstrom:2022yea}.

A different line of research is to treat atomic nuclei and nuclear matter as relativistic systems. In Ref.~\cite{Brockmann:1990cn}, 
with 2NF only, it was shown that the RBHF approach could yield saturation properties closer to the empirical value, in contrast with the non-relativistic BHF theory. Such studies hinted that nuclear matter saturation is a relativistic effect, namely, the Lorentz nature of the scalar and vector interactions leads to density-dependent repulsion, as pointed out in the previous studies~\cite {Walecka:1974qa,Anastasio:1980jm,Walecka:1985my,Serot:1997xg}. In recent years, this line of research has gained much more momentum following the overcoming of several long-standing technical problems~\cite{Brown:1985gt,Ma:2002fm,vanDalen:2004pn,Wang:2021mvg}. However, the two-body bare~\textit{NN}~interaction used in these studies is based on the phenomenological meson exchange picture, i.e., Bonn A, B, and C~\cite{Machleidt:1989tm}, which does not allow a systematic improvement for the description of SNM. From this perspective, a relativistic nuclear force more closely related to the theory of the strong interaction, i.e., Quantum Chromodynamics~(QCD), is needed to gain further insight into the role of relativity in the saturation of nuclear matter. The present work aims to fill this gap.

 Chiral effective field theory~($\chi$EFT), as the low-energy effective theory of QCD, can provide microscopic and QCD-based nucleon-nucleon interactions~\cite{Wein90,Wein91}~needed in~\textit{ab initio}~studies. In the past, most studies focused on non-relativistic nucleon-nucleon interactions~\cite{Epelbaum:2008ga,Machleidt:2011zz}. 
Recently, it was proposed that one can derive a relativistic chiral nucleon-nucleon interaction based on the covariant $\chi$EFT~\cite{Epelbaum:2012ua,Ren:2016jna,Lu:2021gsb}, which provides the much-needed microscopic~\textit{NN}~forces for the RBHF studies~\cite{Shen:2019dls}. The leading order covariant chiral nuclear force can describe the nucleon-proton scattering phase-shits reasonably well~\cite{Ren:2016jna,Ren:2017yvw}, and its renormalizability has been studied, which indeed shows better performance compared to the Weinberg force. For instance, the much-discussed $^3\mathrm{P}_0$ channel is cutoff independent, different from its counterpart in the Weinberg formulation~\cite{Ren:2016jna,Wang:2020myr}.

In this letter, we study the properties of SNM in the RBHF theory.
We employ the relativistic chiral nuclear force to demonstrate how relativity allows one to achieve saturation already with the leading order~(LO)~two-body force, indicating that saturation can indeed be seen as originating from relativistic effects and is undoubtedly a bulk property less affected by the fine details of the nucleon-nucleon interaction. This letter is organized as follows. In Sec.~\ref{sec:level2}, we briefly introduce the theoretical framework of the RBHF theory based on the LO covariant chiral nuclear force. We analyze the results in Sec.~\ref{sec:level3}, followed by a summary and outlook in Sec.~\ref{sec:level4}.

\section{Theoretical framework}\label{sec:level2}

In the RBHF theory, one employs the Dirac equation to describe the single-particle motion of the nucleon in nuclear matter:
\begin{eqnarray}
(\bm{\alpha}\cdot\bm{p}+\beta M+\beta\mathcal{U})u(\bm{p},\lambda)
=E_{\bm{p}}u(\bm{p},\lambda),
\end{eqnarray}
where~$\bm{\alpha}$~and~$\beta$~are the Dirac matrices, $u(\bm{p},\lambda)$~is the Dirac spinor with momentum~$\bm{p}$, single-particle energy~$E_{\bm{p}}$~and helicity~$\lambda$, $M$~is the mass of the free nucleon, and~$\mathcal{U}$~is the single-particle potential operator, providing the primary medium effects. Due to time-reversal invariance, one can neglect the spacelike component of the vector fields, and as a result, the single-particle potential operator can be  expressed as~\cite{Brockmann:1990cn,Tong:2018qwx}:   
\begin{eqnarray}
    \mathcal{U} = U_S + \gamma^0U_0,
\end{eqnarray}
The momentum dependence of the scalar field~($U_S$)~and the timelike component of the vector fields~($U_0$)~is weak~\cite{Gross-Boelting:1998xsk,Wang:2021mvg} and therefore neglected.

The in-medium Dirac equation  can be expressed in the form of the free Dirac equation:
\begin{eqnarray}
    (\bm{\alpha}\cdot\bm{p}+\beta M^\ast)u(\bm{p},\lambda)=E_{\bm{p}}^\ast u(\bm{p},\lambda)\label{equ4},
\end{eqnarray}
by introducing the following effective quantities:
\begin{eqnarray}
    M^\ast = M + U_S,\quad E^\ast_{\bm{p}} = E_{\bm{p}} - U_0,
\end{eqnarray}
whose solution reads
\begin{eqnarray}
    u(\bm{p},\lambda) = \sqrt{\frac{E^\ast_{\bm{p}}+M^\ast}{2M^\ast}}
    \begin{pmatrix}1\\ \dfrac{2\lambda p}{E^\ast_{\bm{p}}+M^\ast}\end{pmatrix}\chi_{\lambda},
\end{eqnarray}
where~$\chi_{\lambda}$~is the Pauli spinor helicity basis. The covariant normalization is $\bar{u}(\bm{p},\lambda)u(\bm{p},\lambda)=1$.

Once $U_S$~and~$U_0$~of the single-particle potential operator are determined, the in-medium Dirac equation can be solved analytically. To achieve this, the matrix element of~$\mathcal{U}$ is constructed following Refs.~\cite{Anas83,Brockmann:1990cn} as
\begin{eqnarray}
    \Sigma(p)
    =\bar{u}(\bm{p},1/2)\mathcal{U}u(\bm{p},1/2)=U_S+\frac{E_{\bm{p}}^\ast}{M^\ast}U_0\label{equ6}
\end{eqnarray}
where the direction of~$\bm{p}$~is taken along the~$z$~axis.

Once~$\Sigma$~is obtained, $U_S$~and~$U_0$~can be determined via the following relations,
\begin{subequations}\label{equ7}
\begin{eqnarray}
\Sigma(p_1) = U_S + \frac{E_{p_1}^\ast}{M^\ast}U_0,\\
\Sigma(p_2) = U_S + \frac{E_{p_2}^\ast}{M^\ast}U_0,
\end{eqnarray}
\end{subequations}
where~$p_1$~and~$p_2$~describe two momenta that are in the Fermi sea.

On the other hand, the matrix elements of~$\mathcal{U}$~in Eq.~\eqref{equ6} can be calculated
as the integrals of the effective~$G$~matrix
\begin{widetext}
\begin{eqnarray}
    \Sigma(p) = \sum_{\lambda'}
    \int_0^{k_F}\frac{\mathrm{d}^3p'}{(2\pi)^3}\frac{M^\ast}{E_{\bm{p}'}^\ast}
    \langle \bar{u}(\bm{p},1/2)
    \bar{u}(\bm{p}',\lambda')
    |\bar{G}|u(\bm{p},1/2)u(\bm{p}',\lambda')\rangle\label{equ8},
\end{eqnarray}
\end{widetext}
where~$\bar{G}$~is the antisymmetrized~$G$~matrix. In the~$no\text{-}sea$~approximation~\cite{Serot:1997xg}, the integral is only performed for the single-particle states in the Fermi sea. 

In the nonrelativistic BHF theory, the Brueckner-Goldstone expansion has been shown to converge in the number of hole lines~\cite{RAJARAMAN:1967jlg} (see also Ref.~\cite{Lu:2017nbi}). The lowest order of the hole-line expansion, the two-hole-line, is the same as the first order $G$-matrix expansion, i.e., the RBHF framework, where the~$G$-matrix is obtained by solving the in-medium relativistic scattering equation. One of the most widely used scattering equations in the RBHF theory is the Thompson equation~\cite{Thompson:1970wt}, a relativistic three-dimensional reduction of the Bethe-Salpeter equation~\cite{Salpeter:1951sz}. To include medium effects, the Thompson equation in the rest frame of nuclear matter reads,
\begin{eqnarray}
    G(\bm{q}',\bm{q}|\bm{P},W)
    =&& V(\bm{q}',\bm{q}|\bm{P})
    +\int\frac{\mathrm{d}^3k}{(2\pi)^3}V(\bm{q}',\bm{k}|\bm{P})\nonumber\\
    &&\times\frac{M^{\ast2}}{E_{\bm{P}+\bm{k}}^\ast E_{\bm{P}-\bm{k}}^\ast}
    \frac{Q(\bm{k},\bm{P})}{W-E_{\bm{P}+\bm{k}}-E_{\bm{P}-\bm{k}}}\nonumber\\
    &&\times G(\bm{k},\bm{q}|\bm{P},W)\label{equ9},
\end{eqnarray}
where~$\bm{P}=\frac{1}{2}(\bm{k}_1+\bm{k}_2)$~is the center-of-mass momentum, and~$\bm{k}=\frac{1}{2}(\bm{k}_1-\bm{k}_2)$~is the relative momentum of the two interacting nucleons with momenta~$\bm{k}_1$~and~$\bm{k}_2$, and ~$\bm{q},\bm{q}'$,~and~$\bm{k}$~are the initial, final, and intermediate relative momenta of the two nucleons scattering in nuclear matter, respectively. $W=E_{\bm{P}+\bm{q}}+E_{\bm{P}-\bm{q}}$~is used to describe the starting energy. $M^\ast$~and~$E^\ast_{\bm{P}\pm\bm{k}}$~are effective masses and energies. The Pauli operator~$Q(\bm{k},\bm{P})$~only allows the scattering of nucleons to un-occupied states, i.e.,
\begin{eqnarray}
    Q(\bm{k},\bm{P})
    =\begin{cases}
    1,\quad |\bm{P}+\bm{k}|,|\bm{P}-\bm{k}|>k_F\\
    0,\quad\text{otherwise}
    \end{cases},
\end{eqnarray}
where~$k_F$~is the Fermi momentum.

One key feature of the RBHF theory is that the~$V$~in Eq.~\eqref{equ9}~should be a bare~\textit{NN}~interaction of covariant form. This work adopts the LO covariant chiral nuclear force~\cite{Ren:2016jna,Xiao:2018jot}. In this order, the relativistic potential~$V$~is the sum of the contact term and the one-pion-exchange diagram,
\begin{eqnarray}
    V_{\mathrm{LO}} = V_{\mathrm{CTP}} + V_{\mathrm{OPEP}},
\end{eqnarray}
where the contact potential~(CTP)~is
\begin{widetext}
\begin{eqnarray}
    V_{\mathrm{CTP}}(\bm{p}',\bm{p}) 
    =&& C_S\left[\bar{u}(\bm{p}',\lambda'_1)u(\bm{p},\lambda_1)\right]
    \left[\bar{u}(-\bm{p}',\lambda'_2)u(-\bm{p},\lambda_2)\right]\nonumber\\
    &&+ C_V\left[\bar{u} (\bm{p}',\lambda'_1)\gamma_\mu u(\bm{p},\lambda_1)\right]
    \left[\bar{u}(-\bm{p}',\lambda'_2)\gamma^\mu u(-\bm{p},\lambda_2)\right]\nonumber\\
    &&+C_{AV}\left[\bar{u} (\bm{p}',\lambda'_1)\gamma_\mu\gamma_5 u(\bm{p},\lambda_1)\right]
    \left[\bar{u}(-\bm{p}',\lambda'_2)\gamma^\mu\gamma_5 u(-\bm{p},\lambda_2)\right]\nonumber\\
    &&+C_T \left[\bar{u} (\bm{p}',\lambda'_1)\sigma_{\mu\nu} u(\bm{p},\lambda_1)\right]
    \left[\bar{u}(-\bm{p}',\lambda'_2)\sigma^{\mu\nu} u(-\bm{p},\lambda_2)\right],
\end{eqnarray}
where~$C_{S,V,AV,T}$~are low-energy constants~(LECs)~to be determined by fitting to the nucleon-nucleon scattering phase shifts. The one-pion-exchange potential~(OPEP)~reads
\begin{eqnarray}
    V_{\mathrm{OPEP}}(\bm{p}',\bm{p})
     = \frac{g_A^2}{4f_\pi^2}\frac{
     \left[\bar{u} (\bm{p}',\lambda'_1)\bm{\tau}_1\gamma^\mu\gamma_5q_\mu u(\bm{p},\lambda_1)\right]
     \left[\bar{u}(-\bm{p}',\lambda'_2)\bm{\tau}_2\gamma^\nu\gamma_5 q_\nu u(-\bm{p},\lambda_2)\right]}{(E_{\bm{p}'}-E_{\bm{p}})^2-(\bm{p}'-\bm{p})^2-m^2_\pi},   
\end{eqnarray}
\end{widetext}
where the pion decay constant~$f_\pi=92.4$~MeV, the axial vector coupling~$g_A=1.29$~\cite{ParticleDataGroup:2016lqr}. $m_\pi$~is the pion mass, $q=(E_{\bm{p}'}-E_{\bm{p}},\bm{p}'-\bm{p})$~represents the four momentum transferred, and~$\bm{\tau}_1,\bm{\tau}_2$~are the isospin Pauli matrix.

Furthermore, the potential has to be regularized to avoid ultraviolet divergences and facilitate numerical calculations. Here, we choose the commonly used separable cutoff function~\cite{Epelbaum:1999dj,Ren:2017yvw}: 
\begin{eqnarray}
    V_{\mathrm{LO}}
    \rightarrow V_{\mathrm{LO}}^{\mathrm{Reg.}}=V_{\mathrm{LO}}\exp\left(\frac{-\bm{p}^{2n}-\bm{p}^{\prime2n}}{\Lambda^{2n}}\right)\label{equ14},
\end{eqnarray}
where~$n=2$~and~$\Lambda$~is the cutoff momentum.
Eqs.~\eqref{equ4},\eqref{equ7},\eqref{equ8}~, and~\eqref{equ9} constitute a set of coupled equations that need to be solved self-consistently. Starting from chosen initial values of~$U_S^{(0)},U_0^{(0)}$, one solves the in-medium Dirac equation~\eqref{equ4} to obtain the Dirac spinors. Next, one solves the Thompson equation~\eqref{equ9} to obtain the~$G$~matrix and uses the integrals in Eq.~\eqref{equ8} to get~$\Sigma$. With Eq.~\eqref{equ7}, one obtains a new set of fields, $U_S^{(1)},U_0^{(1)}$, to be used in the next iteration.    

When $U_S$~and~$U_0$~of the single-particle potential converge, the binding energy per nucleon in SNM can be calculated as
\begin{eqnarray}
    E/A &&=E_T/A + E_V/A\nonumber\\
    E_T/A&&=\frac{1}{\rho}
    \sum_{\lambda}\int_0^{k_F}
    \frac{\mathrm{d}^3p}{(2\pi)^3}
    \frac{M^\ast}{E_{\bm{p}}^\ast}\langle\bar{u}(\bm{p},\lambda)|\bm{\gamma}\cdot\bm{p}+M|u(\bm{p},\lambda)\rangle-M\nonumber\\
    E_V/A&&=\frac{1}{2\rho}
    \sum_{\lambda,\lambda'}
\int_0^{k_F}\frac{\mathrm{d}^3p}{(2\pi)^3}\int_0^{k_F}
    \frac{\mathrm{d}^3p'}{(2\pi)^3}\frac{M^\ast}{E_{\bm{p}}^\ast}\frac{M^\ast}{E_{\bm{p}'}^\ast}\nonumber\\
    &&\times\langle \bar{u}(\bm{p},\lambda)\bar{u}(\bm{p}',\lambda')|\bar{G}(W)|u(\bm{p},\lambda)u(\bm{p}',\lambda')\rangle,\label{equ15}\nonumber\\
\end{eqnarray}
where the isospin indexes are suppressed. The starting energy is ~$W=E_p+E_{p'}$. The density~$\rho$~is related to the Fermi momentum~$k_F$~through~$\rho=2k_F^3/3\pi^2$.

\section{\label{sec:level3} Results and discussions}
The unknown LECs are determined in Refs.~\cite{Ren:2016jna,Lu:2021gsb} by performing a simultaneous fit to the~$J\leq1$~Nijmegen partial wave phase shifts of the~$np$~channel up to the laboratory kinetic energy~($E_{\mathrm{lab}}$)~of 100 MeV at six energies~\cite{Stoks:1993tb}. In the fitting process,
the $\tilde{\chi}^2=\sum\limits_i(\delta^i-\delta^i_{\mathrm{PWA93}})^2$~\cite{Ren:2016jna} is minimized. In Refs.~\cite{Ren:2016jna,Lu:2021gsb}, a sharp cutoff is used. In the present work, we adopt the exponential regulator in Eq.~\eqref{equ14}, which is more amenable for  RBHF studies. In addition, to study the impact of the cutoff on the results, we vary $\Lambda$ from 450 MeV to 600 MeV. The corresponding LECs~ $C_{S,V,AV,T}$~are listed in Table I. In the following, we employ the four LO covariant chiral nuclear forces listed in Table I to calculate phase shifts and properties of SNM.
\begin{table}[h]
\caption{\label{tab:table4}Values of the LO LECs~(in units of $\mathrm{GeV}^{-2}$)~with the cutoff ranging from 450 MeV to 600 MeV.}
\begin{ruledtabular}
\begin{tabular}{cccccc}
$\Lambda(\mathrm{MeV})$&
$C_S$&
$C_V$&
$C_{AV}$&
$C_T$&
$\tilde{\chi}^2/N$\\
\hline
450&-497.51&422.09&-122.29&-54.24&7.43\\
550&-561.94&532.56&-164.86&-68.12&4.66\\
567&-578.05&560.43&-174.54&-71.32&4.25\\
600&-619.06&631.96&-198.56&-79.55&3.57\\
\end{tabular}
\end{ruledtabular}
\end{table}

\begin{figure}[h]
\includegraphics[scale=0.40]{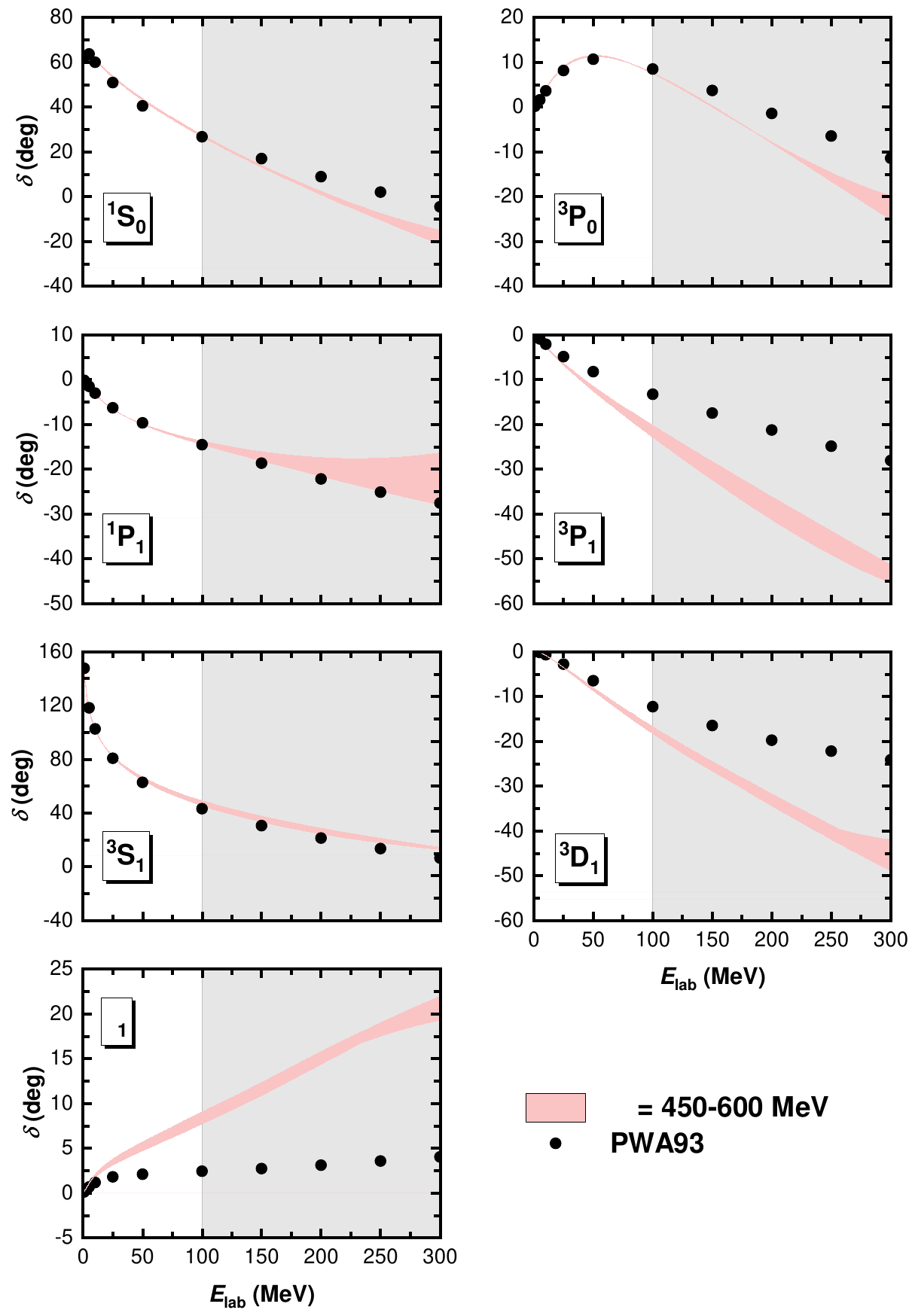}
\caption{\label{fig1}~(Color online)~Neutron-proton phase shifts for partial waves of $J\leq1$. The red bands are those of the LO covariant chiral nuclear force, with the cutoff ranging from 450 MeV to 600 MeV. The solid dotted lines represent the~$np$~phase shift analyses of Nijmegen~\cite{Stoks:1993tb}. The gray backgrounds denote the energy regions where the theoretical results are predictions.}
\end{figure}
With the four sets of LECs, 
the descriptions of the Nijmegen multi-energy $np$ phase shifts up to~$E_{\mathrm{lab}}=300$~MeV are shown in Fig.~\ref{fig1}. The red bands cover the variations from the best-fit results obtained with the cutoff ranging from 450 MeV to 600 MeV. The LO covariant NN force can provide a reasonable description of the $np$ phase shifts of~${}^1\mathrm{S}_0, {}^3\mathrm{P}_0, {}^1\mathrm{P}_1$~and ${}^{3}\mathrm{S}_1$, but the description of the phase shifts of ${}^3\mathrm{P}_1,{}^3\mathrm{D}_1,\varepsilon_1$ is not very satisfactory at higher energies. Furthermore, the cutoff variation from 450 MeV to 600 MeV does not qualitatively change the overall picture.

\begin{figure}[ht]
\includegraphics[scale=0.4]{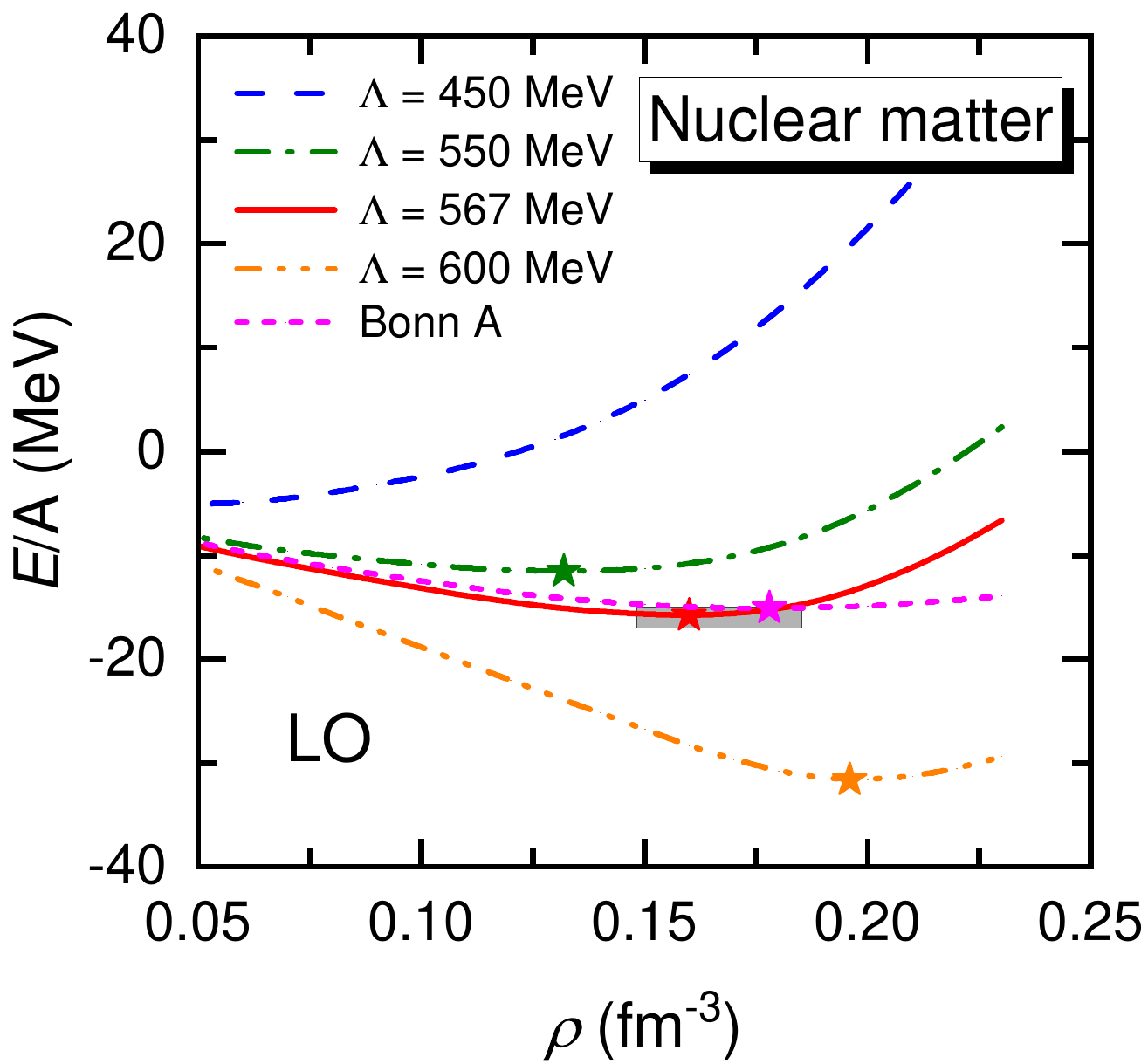}
\caption{\label{fig2}~(Color online)~Energy per nucleon~($E/A$)~in SNM as a function of the density~$\rho$~in the RBHF theory obtained with the leading-order covariant chiral nuclear force\cite{Ren:2016jna}, in comparison with the results obtained with the Bonn A potential~(magenta dash line)~\cite{Brockmann:1990cn}. The pentagrams denote the saturation point. The shaded area indicates the empirical values~\cite{Beth71,Spru72}. }
\end{figure}

In Fig.~\ref{fig2}, we show the corresponding
equations of state (EOS) for SNM. We note that the EOSs become more and more repulsive as the cutoff~$\Lambda$~decreases. At low densities, we observe a weak cutoff dependence. In contrast, at higher densities, the energy per particle of SNM strongly depends on the momentum cutoff~$\Lambda$. We realized in retrospect that the cutoff dependence of the potential could be exacerbated in nuclear matter compared to that in vacuum, similar to the nonrelativistic case~\cite{Sammarruca:2014zia}. The origin of such an exacerbation is not clear currently.  In the future, we will investigate
whether full cutoff independence can be achieved in the relativistic framework.
We further note that for~$\Lambda$~of~$567$~MeV, the binding energy per nucleon obtained in the RBHF theory is~$-15.82$~MeV, in good agreement with the empirical value of~$-16\pm1$~ MeV, and the saturation density~$\rho_0$~is~$0.16$ fm${}^{-3}$~which is also in good agreement with the empirical one of~$0.16\pm0.01$ fm${}^{-3}$. For $\Lambda=567$~MeV, the compression modulus $K_\infty$ at the saturation density is 252 MeV, again in agreement with the empirical value of $240 \pm 20$~MeV~\cite{Garg:2018uam}.
 We expect that the dependence of the results on the cutoff will be alleviated when higher-order relativistic chiral forces are employed.

On the other hand, compared with the results obtained with the Bonn A potential, at densities~$\rho>\rho_0$, the EOSs of SNM are much stiffer, although they agree with each other at lower densities. This might be traced to the fact that the LO covariant chiral nuclear force provides a relatively poor description of the phase shifts of~$P$~partial waves.  

\begin{figure}[ht]
\includegraphics[scale=0.45]{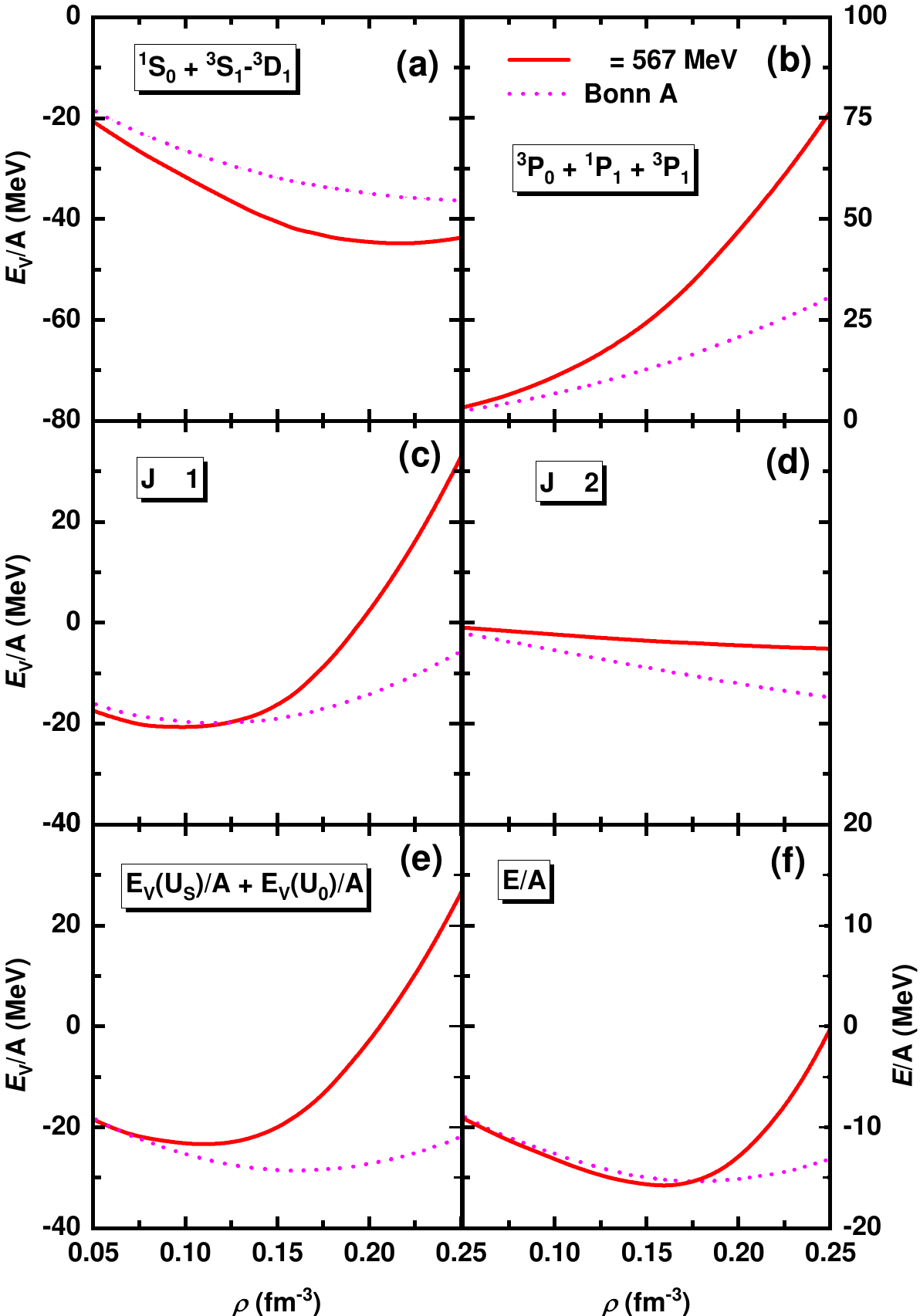}
\caption{\label{fig3}~(Color online)~Partial wave and the scalar and vector potentials' contributions to the potential energy in SNM obtained with the LO chiral nuclear force, in comparison with the results obtained with the Bonn A potential~(magenta dash line). (a): shows the contributions of the~${}^1\mathrm{S}_0$~and~${}^3\mathrm{S}_1$-$
{}^3\mathrm{D}_1$~channels; (b): shows the contributions of P-waves; (c): shows the sum of the contributions of the partial waves with $J\leq1$; (d): shows the sum of the contributions of the partial waves with $J\geq2$; (e): shows the contributions of the scalar and vector potentials; (f): shows the energy per nucleon in SNM.}
\end{figure}

To better understand the EOSs of SNM, we show the contributions of different partial waves to the potential energy and those of the scalar and vector potentials to the potential energy in Fig.~\ref{fig4}. Because the leading-order covariant chiral nuclear force is only fitted to partial waves with $J\leq1$, we only show the contribution of partial waves with $J\leq1$ to the nucleon potential energy. The contribution of higher partial waves with $J\geq2$ to the potential energy is small compared to that of the partial waves with $J\leq1$. However, the contributions of all the partial waves with $J\leq 7$ are considered in obtaining the equation of state. As can be seen from Fig.~\ref{fig3}, for both the chiral force and the Bonn A potential, the~${}^1\mathrm{S}_0$~and~${}^3\mathrm{S}_1$-$
{}^3\mathrm{D}_1$~channels provide attraction and the $P$-waves provide repulsion. Although their magnitude and density dependence differ, the sums of the attraction and repulsion to the potential energy are similar below the saturation density. 
One can conclude that the saturation mechanism of the LO covariant chiral nuclear force is similar to but different from that of the Bonn A potential. This may explain why the EOSs of SNM obtained with the LO covariant chiral nuclear force are much stiffer at densities~$\rho>\rho_0$.

In the RBHF theory, it is known that the larger the scalar potential $U_S$, the larger the relativistic effect~\cite{Shen:2019dls}. In other words, the small component of the Dirac spinor can become
non-negligible even in the low-velocity limit, as $M^\ast$ can be significantly smaller than the bare mass $M$. In Fig.~\ref{fig5}, we compare the full RBHF results and the one obtained by replacing the effective mass $M^\ast$ with the bare nucleon mass $M$ in iteratively solving the set of coupled equations in the RBHF theory. The latter can be viewed as the non-relativistic result~\cite{Brockmann:1990cn}. The results show that the relativistic effect provides repulsion, which is consistent with the results obtained with the Bonn A potential. On the other hand, the density dependencies are different. However, due to the exploratory nature of the LO covariant chiral force employed, we do not discuss further the differences between the results obtained with the Bonn A potential and those obtained with the LO covariant chiral force.    

\begin{figure*}[h]
\includegraphics[scale=0.60]{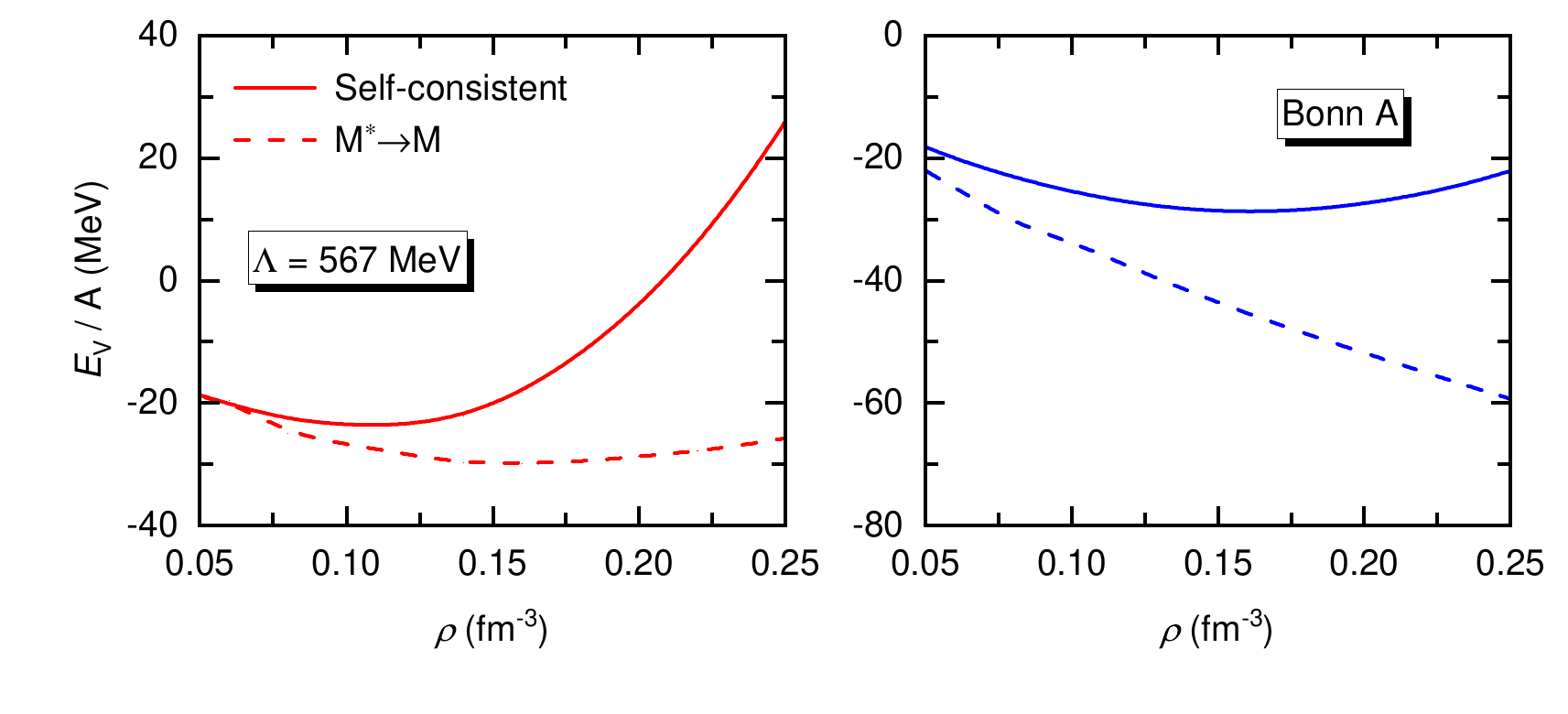}
\caption{\label{fig5}Repulsive relativistic effect on the potential energy of SNM in the RBHF obtained with the LO covariant chiral force, in comparison with the results obtained  with the Bonn A potential.}
\end{figure*}

Next, we compare in Fig.~\ref{fig4} the RBHF results with the BHF results obtained with the non-relativistic LO, NLO, and N$^2$LO chiral nuclear forces~\cite{Sammarruca:2021bpn}. The solid line shows the EOS obtained with the LO covariant chiral nuclear force and a ~$\Lambda = 567$~MeV cutoff. There is no saturation in the BHF results up to NLO. In Refs.~\cite{Sammarruca:2014zia,Sammarruca:2021bpn,Ren:2016jna}, it is shown that the NLO non-relativistic chiral nuclear force can provide a reasonable description of the $np$ phase shifts of partial waves of $J\leq 1$ similar to the LO covariant chiral nuclear force. On the other hand, the NLO non-relativistic chiral nuclear force cannot provide nuclear matter saturation in the BHF framework and all other non-relativistic frameworks.  Only at N${}^2$LO, when 3NF is considered, a reasonable description of nuclear matter saturation can be achieved.  On the other hand, the RBHF theory with the LO covariant chiral nuclear force is sufficient to provide a decent description of the saturation properties of SNM, which indicates that nuclear matter saturation can indeed be understood as a relativistic effect~\cite{Anastasio:1980jm,Brockmann:1990cn} and is independent of the fine details of the nucleon-nucleon interaction. 
Although the $P$ wave contributions to the EOS are pretty different for the LO covariant chiral nuclear force and the Bonn A potential, nuclear matter saturation can be achieved with both forces and without explicitly considering 3NF.

\begin{figure}[ht]
\includegraphics[scale=0.4]{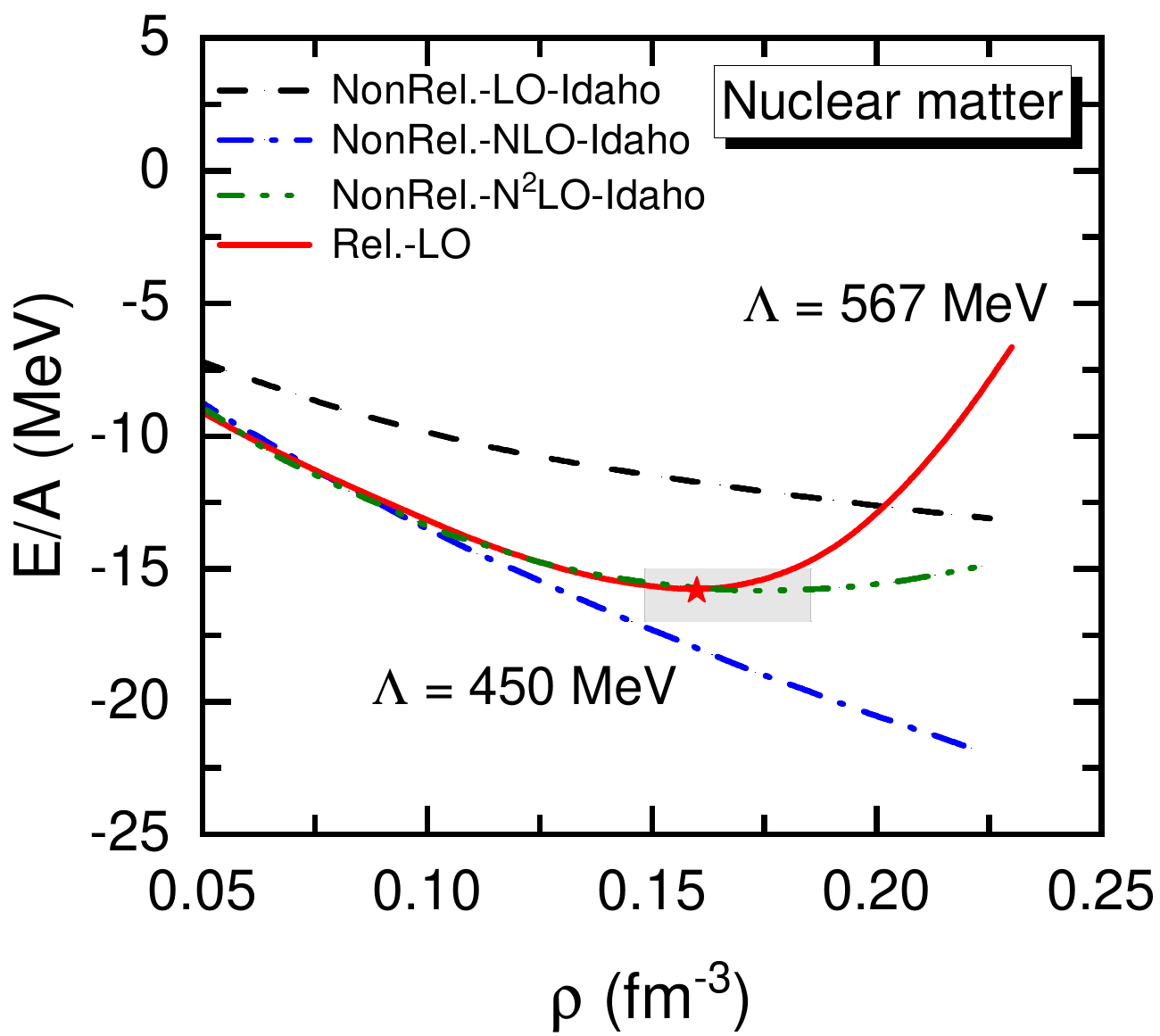}
\caption{\label{fig4}~(Color online)~Energy per nucleon~($E/A$)~in SNM as a function of the density~$\rho$~with the LO relativistic chiral nuclear force~(red solid line), in comparison with the non-relativistic LO, NLO, and N${}^2$LO chiral nuclear force~(black dashed line, blue dotted dash and green dot-dot-dashed)~\cite{Sammarruca:2021bpn}. The shaded area indicates the empirical values~\cite{Beth71,Spru72}.}
\end{figure}

\section{\label{sec:level4} Summary and Outlook}

In summary, we have studied the equation of state in the relativistic Brueckner-Hartree-Fock theory with the LO covariant chiral nuclear force. We found that the saturation properties of symmetric nuclear matter can be reasonably described with a cutoff of ~$\Lambda=567$~MeV. This contrasts with the nonrelativistic Brueckner-Hartree-Fock theory, where reasonable saturation can only be achieved at N${}^2$LO (with 3NF). On the other hand, in the RBHF theory, the saturation properties of symmetric nuclear matter can be reasonably described already at leading order, indicating that nuclear matter saturation can indeed be viewed as a relativistic effect, reinforcing the long-held belief~\cite{Anastasio:1980jm,Brown:1985gt,Brockmann:1990cn,Ma:2002fm,vanDalen:2004pn,Wang:2021mvg}. In addition, we can achieve a simultaneous description of the $np$ phase shifts and the saturation properties of symmetric nuclear matter at leading order without explicitly introducing three-body forces. In this respect, the results of the present work are consistent with the recent works seeking the essential elements of nuclear binding~\cite{Lu:2018bat,Gnech:2023prs}, but in a relativistic framework and highlights the unique role played by consistently treating relativistic degrees of freedom. We stress that this is the first RBHF study employing (covariant) chiral nuclear forces, narrowing the gap between RBHF studies and their non-relativistic counterparts. On the other hand, it should be noted that the dependence of the RBHF results on the momentum cutoff is much stronger than the bare relativistic chiral nuclear force. It is, therefore, desirable to check whether such a dependence becomes milder at higher chiral orders. Indeed, our preliminary results support such a trend.

In the future, we would like to perform studies with higher-order covariant chiral nuclear forces and check the convergence of our results. In addition, we would like to extend such studies to pure neutron matter, whose EOS plays an essential role in heavy-ion physics as well as in models of neutron stars,
gravitational collapse supernovae, and neutron star mergers.

\section*{Acknowledgments}

Wei-Jiang Zou thanks Xiu-lei Ren, Si-bo Wang, and Yi-long Yang for the helpful discussions. We thank Shi-Hang Shen for the enlightening communications. This work was supported in part by the National Natural Science Foundation of China under Grants No.11935003, No.11975031, No.12141501, No.12070131001, No.11735003, No.11975041, No.11961141004, the National Key R\&D Program of China under Grant No.2023YFA1606700, and the State Key Laboratory of Nuclear Physics and Technology, Peking University No.NPT2023ZX03.

\bibliography{apssamp}

\newpage

\appendix

\begin{widetext}
    
\section*{Supplementary Material for “Saturation of nuclear matter in the relativistic Brueckner Hatree-Fock approach with a leading order covariant chiral nuclear force”}

\section*{}
 This supplemental material provides further details useful for understanding the results presented in the main text.
 \section*{The potentials of symmetric nuclear matter}
The scalar and vector potentials are two important quantities in the RBHF model to connect the Dirac equation and G matrices through the nucleon single-particle potential, denoting the attraction and repulsion parts of the NN interaction at different ranges. We present in Fig.\ref{fig3} the
momentum cutoff dependence of the single-particle potentials $U_S$ and $U_0$. Both
 attractive $U_S$ and repulsive $U_0$, as large as several hundred MeV, are generated for all cutoffs. It can be seen from Fig.~\ref{fig3} that the sensitivity on the momentum cutoff mainly comes from the Lorentz vector potential. The results hinted
that nuclear matter saturation is a relativistic effect, namely,
the Lorentz nature of the scalar and vector interactions leads
to density-dependent repulsion, as pointed out in the previous studies\cite{Walecka:1974qa,Anastasio:1980jm,Walecka:1985my,Serot:1997xg}.
\begin{figure*}[h]
\includegraphics[scale=0.35]{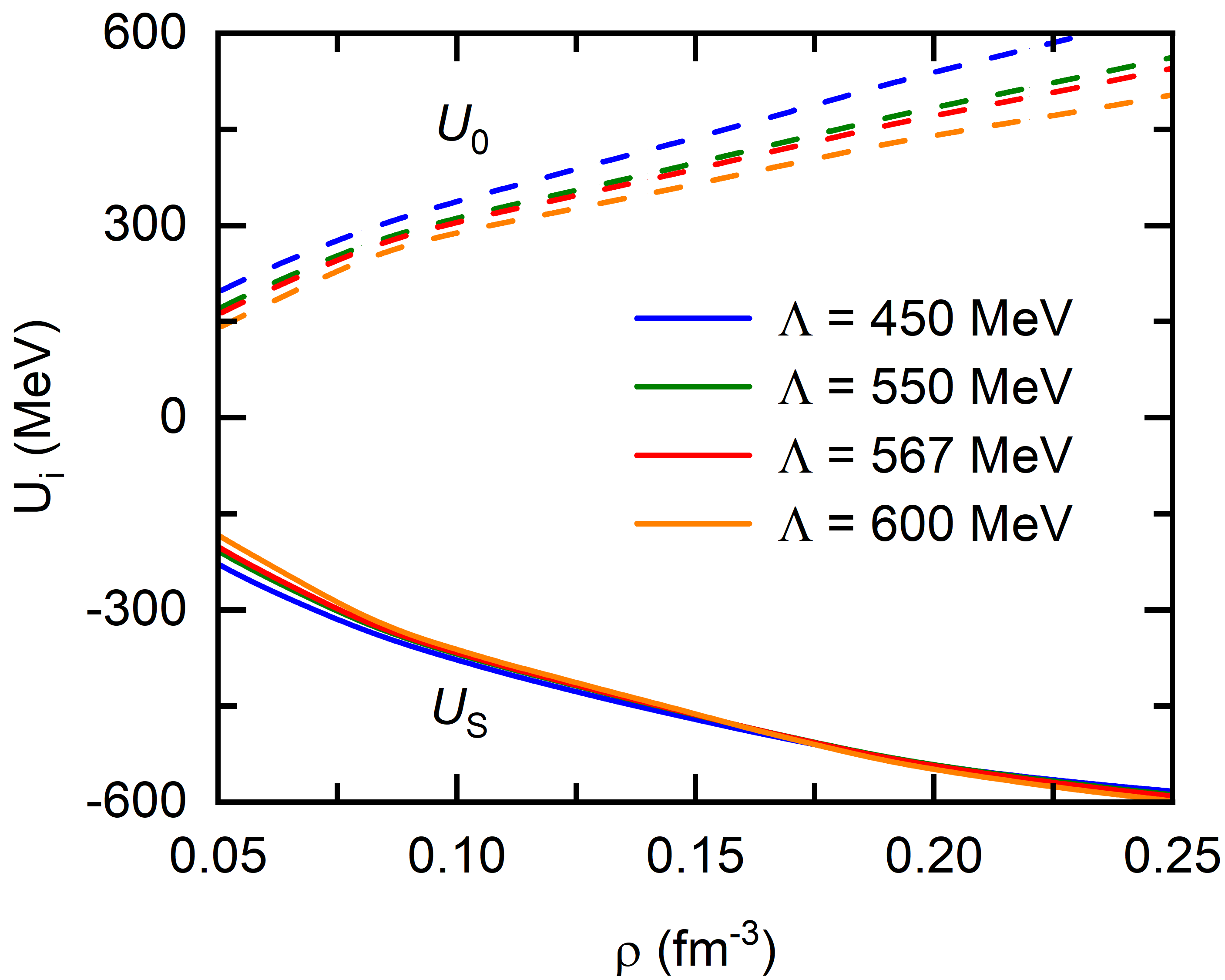}
\caption{\label{fig3} (Color online) Momentum cutoff dependence of the single-particle scalar~$U_S$~and vector~$U_0$ potentials.  }
\end{figure*} 

\section*{Partial wave contributions to the potential energy }
It is instructive to decompose the partial wave contributions to the potential energy and compare them with the
results obtained with the Bonn A potential.
As can be seen from Fig.~\ref{fig5},  
for both the chiral force and the Bonn A potential, the~${}^1\mathrm{S}_0$~and~${}^3\mathrm{S}_1$-$
{}^3\mathrm{D}_1$~channels provide attraction, and the $P$-waves provide repulsive contributions. Although their magnitude and dependence on density differ, the sums of the attraction and repulsion to the potential energy are similar below the saturation density. 
One can conclude that the saturation mechanism for the leading-order (LO) covariant chiral nuclear force is similar but different from that for the Bonn A potential. One should not over-interpret the difference because the chiral force is only at leading order while the Bonn A potential is more accurate.

\begin{figure*}[h]
\includegraphics[scale=0.45]{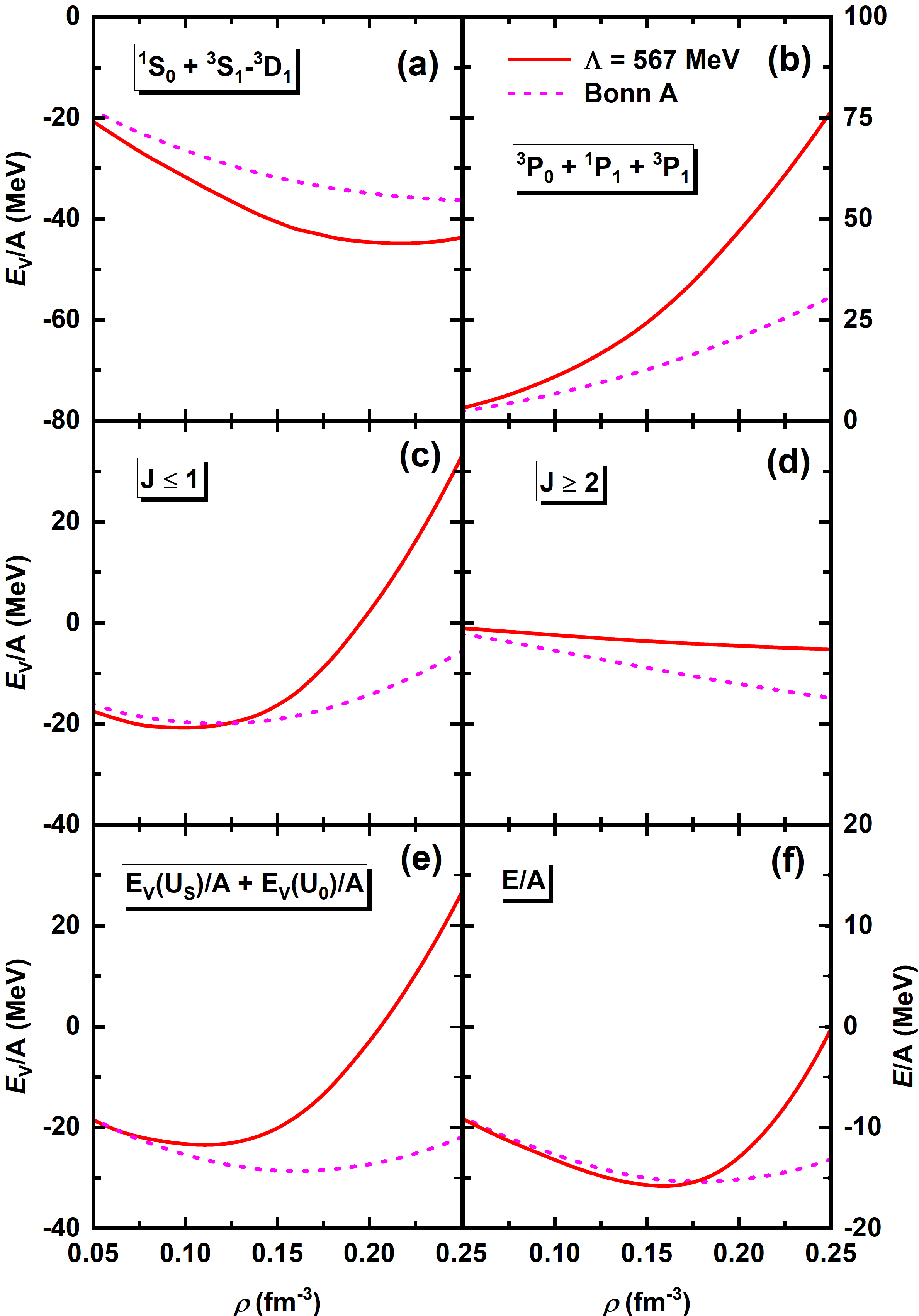}
\caption{\label{fig5}~(Color online)~Partial wave and the scalar and vector potentials' contributions to the potential energy in SNM obtained with the LO chiral nuclear force, in comparison with the results obtained with the Bonn A potential. (a): shows the contributions of the~${}^1\mathrm{S}_0$~and~${}^3\mathrm{S}_1$-$
{}^3\mathrm{D}_1$~channels; (b): shows the contributions of $P$-waves; (c): shows the sum of the contributions of the partial waves with $J\leq1$; (d): shows the sum of the contributions of the partial waves with $J\geq2$; (e): shows the contributions of the scalar and vector potentials; (f): shows the energy per nucleon in SNM.}
\end{figure*}

As seen from Fig.~\ref{fig5}, the contribution of the partial waves $J\geq2$ to the potential energy is very small compared to the contribution of the partial waves $J\leq1$.

We show the contribution of ${}^3\mathrm{P}_2$, ${}^1\mathrm{D}_2$, and ${}^3\mathrm{D}_2$  in Fig.~\ref{fig1}. We note that the contributions of ${}^3\mathrm{P}_2$, ${}^1\mathrm{D}_2$, and ${}^3\mathrm{D}_2$ depend weakly on the momentum cutoff.
In addition, they are all smaller than those of the corresponding Bonn A potential. Given that the contributions of  $P$-waves  are more repulsive than those of the Bonn A, in the covariant chiral force we do not have the cancellations observed in the Bonn A potential, as shown in Fig.~\ref{fig6}. This is unsurprising because
the leading order covariant chiral force differs significantly from the Bonn A potential. Nonetheless, the sum of all the partial waves is similar to that of the Bonn A potential below the saturation point, as shown in Fig.~\ref{fig5}. 

\begin{figure*}[h]
\includegraphics[scale=0.25]{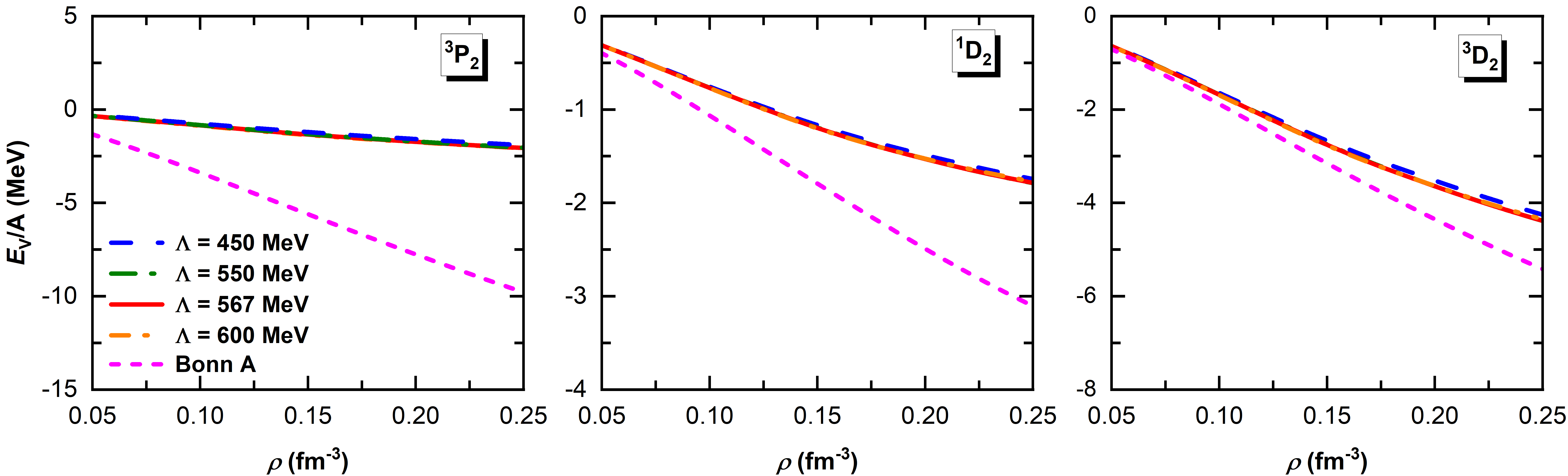}
\caption{\label{fig1} Contributions of ${}^3\mathrm{P}_2$, ${}^1\mathrm{D}_2$, and ${}^3\mathrm{D}_2$  to the potential energy in SNM with the LO chiral nuclear force, in comparison with the results obtained with the Bonn A potential.}
\end{figure*}

\begin{figure*}[h]
\includegraphics[scale=0.35]{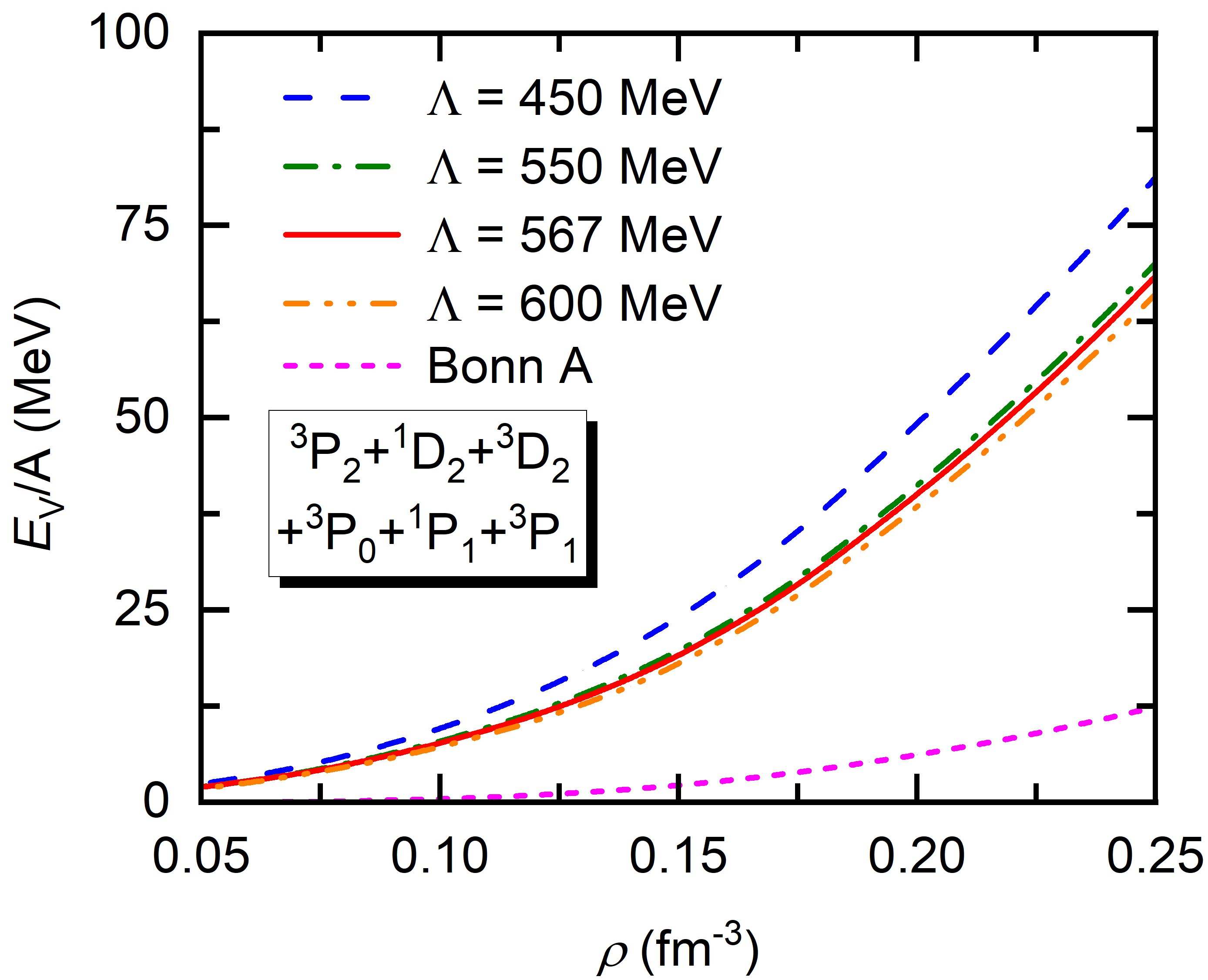}
\caption{\label{fig6} Contributions of ${}^3\mathrm{P}_2$+${}^1\mathrm{D}_2$+${}^3\mathrm{D}_2$+${}^3\mathrm{P}_0$+${}^1\mathrm{P}_1$+${}^3\mathrm{P}_1$ to the potential energy in SNM obtained with the LO chiral nuclear force, in comparison with the results obtained with the Bonn A potential.}
\end{figure*}

We also show the Lorentz
scalar and vector contributions to the potential energy in Fig.~\ref{fig2}

\begin{figure*}[h]
\includegraphics[scale=0.35]{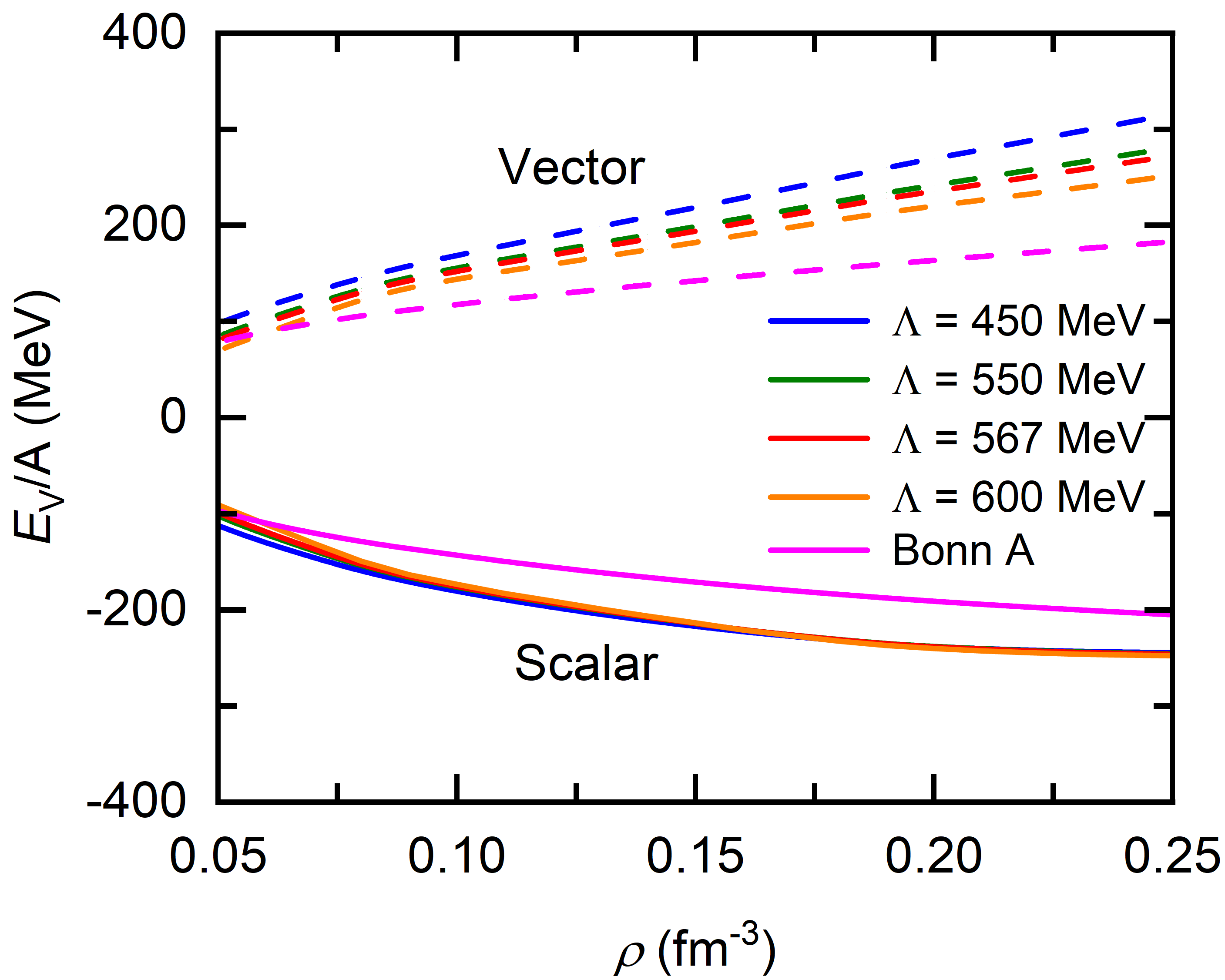}
\caption{\label{fig2}Contributions of the
scalar and vector potentials, $U_S$ and $U_0$, to the potential energy in SNM, in comparison with the corresponding results obtained with the Bonn A potential.}
\end{figure*}

From Fig.~\ref{fig2}, one can see that the contribution of the scalar and vector potentials, $U_S$ and $U_0$, to the potential energy, are similar to those obtained with the Bonn A potential.


\end{widetext}

\end{document}